# Source Attribution and Audience Uptake in Digital Witnessing: Evidence from a Chinese Diaspora Account on X

Huiqian Lai
Syracuse University
hlai12@syr.edu

Runze Ding
Swansea University
runze.ding@swansea.ac.uk

## Abstract

*Digital-witnessing research explains how citizen testimony becomes publicly visible. Less is known about audience uptake after publication and whether the attributed source corresponds to the form of response. This study analyzes 47,635 posts (April 2022 to May 2026) from Teacher Li Is Not Your Teacher, a diaspora-based account that curates and publishes reports and testimony from inside China for transnational audiences. Among 25,093 witness-relevant posts, 3,601 explicitly attribute their material to ordinary submitters. These posts receive more replies than other witness-relevant posts across adjusted analyses and show a more reply-oriented profile. The primary doubly robust retweet estimate is near zero. Among audited replies to November 2022 user-submitted posts, 91.4% substantively engage the focal material. The pattern is consistent with ordinary-person source attribution presenting testimony as situated and open to public response. The study extends digital witnessing beyond visibility by separating redistribution from the public response that gathers around testimony.*

**Keywords:** digital witnessing, source attribution, audience uptake, diaspora media, China

## 1. Introduction

Teacher Li Is Not Your Teacher[1] (hereafter, "Teacher Li" or "Li") is a prominent Chinese-language account on X, operated from Italy by Li Ying, that curates and publishes material documenting events inside China. As of August 2026, its public profile displayed approximately 2.2 million followers (Teacher Li Is Not Your Teacher, n.d.). The account became widely known during the 2022 White Paper protests, when it served as a central hub for photographs, videos, and eyewitness reports from inside China (Yang, 2022). Its position outside mainland China's domestic platform environment enables Li to carry locally produced material into transnational public view, often when the underlying events receive little domestic coverage or when related content is removed from Chinese platforms (Zeng & Cheng, 2025). Li first developed this publishing practice on Weibo, and repeated account closures led him to move it to Twitter in April 2022 (Teacher Li Is Not Your Teacher, 2022). On X, the account foregrounds protests, labor disputes, accidents, censorship, coercive governance, and citizen grievances, while also publishing scenes of everyday life and Li's political commentary (Y. Wu, 2025). Selecting and publishing reports about these events is itself a mode of intervention in Chinese public affairs from abroad (Zeng & Cheng, 2025). The account's witness function lies in this editorial selection: Li identifies underreported events and grievances, decides which warrant wider attention, and brings them into public view.

Witness material reaches the account through several publication routes. Ordinary users can send photographs, videos, testimony, and local observations through direct messages (Kuo & Wu, 2022; Y. Wu, 2025). Li reviews these submissions, checks details against other material or public information, writes or edits captions, and publishes selected items under an explicit submission marker.[2] He also publishes material located on Chinese-language platforms or news sources, attributes reports to media organizations, institutions, and official bodies, and publishes commentary, replies, and operational updates in his own voice. User-submitted posts reflect two stages of selection. Submitters decide which events, evidence, and grievances warrant exposure through Teacher Li; Li decides which materials enter the account and how they are captioned, contextualized, and updated. In this study, a *user-submitted witness post* is a post in which Teacher Li explicitly attributes the underlying photograph, video, testimony, or local observation to an ordinary submitter. Witness posts based on other sources draw on material located online or attributed to media, institutions, or official bodies. Posts in Li's own voice enter the witness-relevant corpus when they contain event documentation, testimony, visual evidence, or closely related witness material.

Teacher Li's publishing practice provides infrastructure for transnational Chinese-language counterpublic communication (Fraser, 1990; Warner, 2002; Zeng & Cheng, 2025). Digital witnessing depends

This paper has been accepted for the upcoming 60th Hawaii International Conference on System Sciences (HICSS-60).

on intermediaries who gather fragmented accounts, assess their provenance, and make them legible beyond the settings in which events occur (Chouliaraki, 2015; Mortensen, 2015), and diasporic citizen journalism carries this work across national and platform boundaries (Zeng & Cheng, 2025). Repeated publication under a stable account gives dispersed Chinese-speaking users a recurring site organized around claims marginalized in domestic information channels.

The account's counterpublic role extends into the ways audiences take up witness material after publication. Digital-witnessing research explains how citizen testimony is recorded, mediated, circulated, and made available for interpretation or verification (Chouliaraki, 2015; Zubiaga et al., 2016). Source-cue research links attribution to credibility judgments, content selection, and sharing intentions (Kaiser et al., 2021; Metzger et al., 2010; Yaqub et al., 2020). These literatures say less about whether displayed provenance within a curator-operated witness account corresponds to distinct forms of audience response. Retweets and replies make two response pathways observable: retweets redistribute a published claim to additional networks, while replies attach a textual response to the claim and create a public thread around it (boyd et al., 2010; Honeycutt & Herring, 2009). Explicit ordinary-person attribution may position a post as situated testimony whose context and relation to other evidence invite supplementation, evaluation, and contestation through replies. The study asks:

RQ: Among witness-relevant posts published by Teacher Li on X, how is explicit ordinary-person source attribution associated with reply volume, retweet volume, and the relative balance between replies and retweets?

The dataset covers April 2022, when Li's active publication began, through May 2026, when data collection closed (Teacher Li Is Not Your Teacher, 2022), and contains 47,635 publicly retrievable posts. Computational coding validated against human judgments identifies 25,093 witness-relevant posts, including 3,601 with explicit ordinary-person submission markers. Across regression, matching, and weighting estimators, user-submitted witness posts are associated with higher reply volume than other witness-relevant posts, while the primary doubly robust retweet estimate is near zero. Reply share and the reply–retweet log ratio are also higher for user submissions in the full witness-relevant sample, with less uniform evidence in a stricter formal-source comparison. Semantic-content matching retains a smaller positive reply difference among closely matchable posts. To check reply content, a manual audit examined 72 user-submitted witness posts from November 2022 with at least five retrievable replies and sampled five replies from each. Of the 360 audited replies, 329, or 91.4%, engaged the focal witness material. The source category combines ordinary-person provenance, Li's editorial selection, and displayed attribution; the estimates describe this combined publishing form.

The study makes two contributions. First, it extends digital-witnessing and source-cue research from the visibility and credibility of testimony to audience uptake after publication. The reply-oriented pattern documented here is consistent with ordinary-person source attribution presenting selected testimony as situated and open to continued public engagement. Second, separating replies from retweets specifies two interaction pathways through which a diaspora-based witness account supports counterpublic communication: redistribution across networks and public interpretation within threads. The distinction gives counterpublic analyses of witness accounts an observable trace of each pathway.

## 2. Literature Review and Hypotheses

### 2.1. Chinese Digital Diaspora and Counterpublic Communication on X

China-focused political communication on X unfolds across a divided platform environment. Reports, visual evidence, and testimony often originate within mainland China, where state and platform controls shape political expression, collective-action information, and collaborative verification (King et al., 2013; X. Wu & Fitzgerald, 2021; Zeng et al., 2019). Publication and discussion can then move to X, which is blocked in mainland China and operates outside the direct governance of domestic platforms (Zeng & Fang, 2025). This cross-platform movement separates the information environments of production and uptake.

Digital diaspora research describes networked media as a means through which dispersed populations maintain homeland ties, build political communities, and intervene in public affairs across national borders (Brinkerhoff, 2009). One such field, and the broader social setting for this movement, is the simplified-Chinese sphere on X, often called *Jianzhong Quan* (简中圈), a loosely bounded networked public whose discussions center on Chinese social, cultural, and political affairs (Zeng & Fang, 2025). Its participants include diaspora members, exiled activists, mainland users who cross the Great Firewall, citizen journalists, information accounts, nationalist actors, and state-linked accounts (Zeng & Fang, 2025). Linguistic and issue ties to China coexist with divergent political orientations, producing competition over which events gain visibility and how they are interpreted. Within this heterogeneous field, diasporic media, exile media, and overseas-based influencers with a critical orientation form what Lee and Tse (2025, p. 1) call a "*critical online diasporic infosphere*" that supplies critical information and perspectives to users abroad and in the home country.

Counterpublic theory specifies the political work of this critical diasporic layer. Counterpublics develop through discourse that circulates marginalized accounts, links dispersed participants, and addresses dominant representations and institutions (Fraser, 1990; Warner, 2002). Networked platforms extend these practices across distance and enable dissenting claims to gather around shared events and symbols (Jackson & Welles, 2015). On Chinese-language X, counterpublic communication occupies the critical subset of this heterogeneous field, as its actors contest state-centered accounts and expand the visibility of claims restricted within domestic channels (Lee & Tse, 2025; Zeng & Fang, 2025). Interaction counts provide observable traces of how claims circulate and attract public response; shared political orientation, collective identity, and durable counterpublic formation, however, require evidence from discourse and sustained relations.

Cross-border digital witnessing is one practice through which this counterpublic communication takes form. During the White Paper protests, this sphere supported cross-border news production and collective sense-making (Zeng & Cheng, 2025). Accounts such as Teacher Li collected dispersed reports, selected and contextualized them, and published them for transnational audiences (Kuo & Wu, 2022; Yang, 2022; Zeng & Cheng, 2025). The political setting therefore places curatorial intermediation between situated testimony and audience response.

## 2.2. Digital Witnessing as a Relational and Curated Process

Digital witnessing builds on relational accounts of witnessing in which testimony acquires public force through address, mediation, and reception (Kyriakidou, 2015; Peters, 2001). Testifiers claim situated access to an event and contribute an account, recording, or observation. Intermediaries organize how that material enters public view, and audiences interpret, evaluate, and circulate the published claim. The public significance of witness material rests on the relation among situated testimony, intermediary publication, and audience uptake.

Digital platforms redistribute these relations by allowing participants, bystanders, and local observers to publish witness material outside professional news institutions (Mortensen, 2015). In conflict, crisis, and protest settings, citizen-produced photographs, videos, and accounts may circulate ahead of professional verification and editorial synthesis (Andén-Papadopoulos, 2014; Chouliaraki, 2015). As fragments move beyond the circumstances in which they were produced, their public meaning depends on how other actors select, contextualize, and present them.

Research on gatewatching and curated flows describes actors who monitor distributed sources, select material, and organize it for public attention (Bruns, 2005; Thorson & Wells, 2016), and participatory-journalism research shows that user contributions enter public communication through editors who retain authority over selection, moderation, and framing (Hermida & Thurman, 2008). In digital witnessing, this curatorial work connects scattered material to an intelligible account through verification and contextualization (Chouliaraki, 2015; Hermida, 2012). Diasporic citizen journalism carries this work across platform and national boundaries by gathering material from one information environment and publishing it for audiences in another (Zeng & Cheng, 2025). Each published witness post consequently combines situated source material with an intermediary's decisions about selection and presentation.

Publication opens witness material to further interpretation and verification. Images, recordings, and local accounts may reach audiences with incomplete provenance, missing context, or competing interpretations of what they show (Chouliaraki, 2015). Journalists and networked publics in turn clarify, corroborate, reinterpret, and contest such claims (Hermida, 2012; Zubiaga et al., 2016). Post-publication uptake therefore forms part of the public life of witness material.

## 2.3. Differentiated Forms of Audience Uptake

Audience uptake in digital witnessing consists of the post-publication actions through which platform users respond to published testimony. Users can add local or prior knowledge, corroborate or question details, compare an account with other evidence, interpret the conduct of actors, contest a claim, or carry the material into further networks (Hermida, 2012; Zubiaga et al., 2016). Audience uptake therefore includes both the circulation of a witness claim and the public responses that accumulate around it after publication.

Platform actions make these forms of uptake visible in different ways. Retweeting sends an existing post into additional networks and extends the reach of its claim (boyd et al., 2010). Replying attaches a textual response to the focal post, can build a visible conversational thread around it, and lets dispersed users respond to the same event in view of one another (Honeycutt &

Herring, 2009). Engagement measures therefore carry different meanings depending on the action they record. Because the focal witness post presents a claim about a specific event, reply threads can accumulate local context, corroboration, skepticism, correction, interpretation, and evaluation around a common witness claim (Hermida, 2012; Zubiaga et al., 2016).

Reply volume captures the amount of public textual response attached to a witness post. Its substantive meaning depends on what those responses contain, because public replies may provide information or opinions, express emotion or humor, take an argumentative stance, or move off topic (Zhang et al., 2024). Reply volume is therefore a bounded trace of public response. Source-cue research leaves open whether visible provenance is associated with reply volume or with the balance between replies and retweets.

## 2.4. Source Attribution in Curated Witness Posts

Source information is one visible feature of a published witness post. Credibility research shows that source cues can shape judgments about origin, expertise, and agency (Metzger et al., 2010). Experimental and observational studies also connect source or recommender cues to truth judgments, news selection, and sharing intentions (Kaiser et al., 2021; Nadarevic et al., 2020; Yaqub et al., 2020). The size and direction of these associations vary across source cues and decision contexts (Mang et al., 2024).

Source attribution in a curated witness post identifies the origin of material that has passed through editorial selection and presentation. On Teacher Li's account, a submission marker indicates that the underlying photograph, video, testimony, or local observation comes from an ordinary submitter, while Teacher Li remains the publisher who selects the material and presents it to the account's audience (Kuo & Wu, 2022; Y. Wu, 2025). Audiences therefore encounter ordinary-person provenance and editorial selection together in the same post.

When a witness post attributes its material to an ordinary person, the visible source relation foregrounds situated access and local knowledge. Citizen witness material can derive public relevance from proximity to an event and access unavailable to professional or institutional actors (Andén-Papadopoulos, 2014). Ordinary-person testimony on social media can also draw rhetorical force from personal experience and public address (Kampka & Modrzejewska, 2026). Posts attributed to media organizations, institutions, or official bodies present the material through formal source authority.

In a witnessing context, ordinary-person source attribution may leave more contextual work available to audiences than formal attribution does. A situated account can prompt users with relevant knowledge to add missing details, compare the account with other evidence, corroborate or question what occurred, and evaluate its meaning. Replies provide a direct site for these responses because each reply remains attached to the focal witness post (Honeycutt & Herring, 2009).

Witness-relevant posts carrying ordinary-person source attribution should therefore attract more public textual response than other witness-relevant posts when audiences use replies to supplement, evaluate, or contest situated testimony.

***H1.*** *Ordinary-person source attribution is associated with higher reply volume among witness-relevant posts.*

Retweet volume carries no directional expectation under this reasoning. Ordinary-person source attribution could widen redistribution by marking material as otherwise unavailable, or it could narrow redistribution if audiences pass on formally sourced claims more readily (Yaqub et al., 2020). Redistribution also responds to content features unrelated to source attribution: negative political content spreads faster and farther on Twitter (Antypas et al., 2023), and witness material about harm, coercion, or repression is largely negative. The analysis therefore examines retweet volume as an outcome without a directional hypothesis.

Source attribution may also correspond to the relative balance of audience response. If ordinary-person source attribution is associated with greater public textual response around witness material, replies should account for a larger share of observed reply–retweet activity. This pattern can appear in a higher reply share and a higher reply–retweet log ratio even when retweet volume changes little.

***H2.*** *Ordinary-person source attribution is associated with a more reply-oriented uptake profile among witness-relevant posts, reflected in a higher reply share of reply–retweet activity and a higher reply–retweet log ratio.*

# 3. Data and Methods

The study combined computational content coding, multivariable adjustment, propensity-score matching, semantic-content matching, and manual validation of reply content.

## 3.1. Case Selection and Analysis Window

The Teacher Li account provided a within-account setting in which witness material appeared under ordinary-person, media, institutional, official, and other source attributions, so audience response could be compared across differently attributed material. Replies and retweets provided observable traces of audience uptake; the available data did not identify whether a submitter later engaged with the post. The analysis window ran from April 2022, when Li began actively publishing on the platform, through May 2026, when data collection closed (Teacher Li Is Not Your Teacher, 2022). The window included the November 2022 White Paper protests as one episode within a longer publishing record (Zeng & Cheng, 2025). The estimates therefore apply to posts from this account during this window and do not represent all witness communication on X or in China. Appendix A.1 documents the account and interface context, and Appendix A.2 provides the source-attribution categories, coding definitions, and paraphrased examples.

### 3.2. Data Collection and Analytic Corpus

The dataset contained 47,635 publicly accessible posts published by @whyyoutouzhele during the analysis window: 39,008 standard posts and 8,627 replies. Data collection occurred on May 30–31, 2026, through TwitterAPI.io (TwitterAPI.io, 2026), and records were de-duplicated by post ID. The post-type fields identified no quote posts authored by the account and no pure reposts. Retained fields included post text, publication time, media attachments, and reply, retweet, like, view, and quote-post counts. The provider's public search index excluded posts that were unavailable or not indexed when queried, so the corpus was not treated as a census of every post ever published by the account.

Because engagement measures were cumulative snapshots from that collection, posts published at different times had different periods in which to accumulate responses. All adjusted models therefore included year-month fixed effects, which also absorbed common monthly changes in account reach and platform conditions.

The primary analyses focused on *witness-relevant posts*, defined as posts containing event documentation, testimony, visual evidence, or closely related witness material. A post passed the witness-relevance screen when its visible text established a concrete witnessing object, a public stake, and a witnessing relation; Appendix A.1 provides the full decision rules and boundary cases. Appendix A.3 shows the analytic flow.

### 3.3. Coding Procedure and Validation

An iteratively developed codebook defined witness relevance and source attribution. It specified substantive categories, visible Chinese-language source markers, and decision rules for ambiguous cases. Appendix B contains the full codebook, annotation prompt, marker dictionary, and anonymized examples.

Two human coders independently annotated a 200-post pilot sample. Agreement was 95% for witness relevance and 93% for source type. Disagreements informed revisions to the codebook before computational annotation.

Gemma 4 [3] and Hermes 3 Llama 3.1 [4] then coded the full corpus independently through the DeepInfra API with the finalized prompt. When the two models agreed, their shared label was retained. Two human coders independently reviewed model disagreements and resolved them through adjudication. Inter-model agreement was 88.5% for witness relevance ($\kappa = .814$) and 90.0% for source type ($\kappa = .848$).

The finalized coding procedure was evaluated on a held-out sample of 300 posts independently annotated by the human coders and excluded from codebook development and prompt refinement. Against adjudicated human labels, witness-relevance coding achieved 94.0% accuracy and $\kappa = .87$. Source-type coding achieved 91.3% accuracy, macro-F1 of .88, and $\kappa = .86$.

### 3.4. Source Attribution Variable

The primary explanatory variable was ordinary-person source attribution, operationalized as a binary indicator for user-submitted witness posts. A post was coded as carrying ordinary-person source attribution when Teacher Li explicitly attributed its underlying witness material to an ordinary submitter through the primary submission marker or an equivalent marker. The category [5] contained 3,601 posts, all within the witness-relevant sample. The observed indicator combined ordinary-person provenance, Li's editorial selection, and the displayed attribution. The design did not separately identify the contribution of these components. Appendix A.3 reports sample composition.

Two text-based specifications tested sensitivity to the source-coding rule. The literal-marker specification required the exact primary phrase （网友投稿）. The expanded-marker specification covered additional submission terms and short attribution phrases ending in a submission marker before punctuation. Both were analyzed separately from the source-coded measure.

A focused comparison restricted the reference group to witness-relevant posts attributed to institutional, media, or official sources. The comparison placed ordinary-person source attribution against explicitly formal attribution. It narrowed the comparison group but did not remove differences in post content or editorial selection. Appendix D.3 reports the focused comparison and source-definition sensitivity estimates.

### 3.5. Outcome Variables

The primary outcome was reply volume, measured as log (1 + replies). Replies captured public textual responses attached to a witness post; the manual audit described below assessed whether sampled replies substantively engaged with the focal material. Retweet volume, measured as log (1 + retweets), captured redistribution to additional networks.

Two additional outcomes compared these forms of audience response: reply share, defined as replies divided by the sum of replies and retweets, and the reply–retweet log ratio, defined as log (1 + replies) − log (1 + retweets). Higher values on either measure indicated relatively more reply activity than redistribution. Nineteen posts with neither replies nor retweets were excluded only from reply-share models. Likes and quote-post counts were not used as primary outcomes. Liking neither redistributed a post nor attached a public textual response, and quote-post field coverage was incomplete across the collection period. Appendix C.1 provides full outcome definitions, construction decisions, and supplementary engagement measures.

### 3.6. Manual Validation of Reply Content

To assess whether reply volume captured substantive audience response, a manual audit examined replies to user-submitted witness posts from November 2022. The month covered the height of the White Paper protests, when account activity peaked, and was the period with a separate reply-level corpus. The audit included 72 focal posts with at least five retrievable replies, stratified by total reply volume. Five replies were randomly sampled from each post, yielding 360 replies. Two coders independently classified each reply as substantive uptake or non-substantive activity. Substantive uptake required intelligible engagement with the event, evidence, actors, claims, or meaning presented in the focal post. Appendix C.2 describes reply retrieval, sampling, coding criteria, and intercoder-reliability procedures.

### 3.7. Covariates and Content Robustness

The primary adjusted models accounted for observable differences in publication timing and post format using year-month fixed effects, character count, media count, and six media-format indicators. Timing and format adjustment could not eliminate content differences: user-submitted posts could concern systematically different incidents, evidence, or editorial selections from other witness-relevant posts. All adjusted estimates were therefore interpreted as conditional associations. A content-robustness analysis additionally matched user-submitted witness posts to other witness-relevant posts with similar textual content. Matching used multilingual sentence embeddings, restricted comparisons by publication month and media configuration where possible, and retained only pairs above a cosine-similarity threshold of .70. Appendix D.1 provides text-cleaning rules, matching specifications, thresholds, and balance diagnostics.

### 3.8. Estimation Strategy

The primary estimation sample consisted of witness-relevant posts. Because ordinary-person source attribution was neither assigned nor independently manipulated, the analysis estimated adjusted associations throughout. It first compared outcome distributions using group medians and Mann–Whitney *U* tests, then estimated these associations using ordinary least squares (OLS), propensity-score matching (PSM), and cross-fit augmented inverse probability weighting (AIPW). OLS and AIPW adjusted for the timing and format covariates described above. Cross-fit AIPW served as the primary adjusted specification, while propensity-score matching provided a complementary balance and point-estimate robustness check. Sensitivity analyses restricted the reference group to formal sources, varied the operational definition of user submission, excluded posts published within seven days of the engagement snapshot, and expanded estimation to the full corpus. Appendix D.2 provides estimator specifications and uncertainty calculations, and Appendix D.3 documents the sensitivity analyses.

### 3.9. Ethical and Reproducibility Considerations

The analysis used public posts and publicly visible engagement counts, but the political sensitivity of the documented events created potential risks for people represented in or responding to the material. The study involved no interaction with users and collected no private account data. This analysis of publicly available posts and aggregate measures did not require institutional review. The online supplementary materials are provided as Appendix A, Appendix B, Appendix C, and Appendix D.

## 4. Findings

### 4.1. Analytic Sample

The corpus of 47,635 posts contains 25,093 witness-relevant posts, defined as posts with event

**Table 1. Main outcome contrasts between user-submitted and other witness posts**

| Design | Reply volume log (1 + replies) | Retweet volume log (1 + retweets) | Reply share | Reply-retweet log ratio |
|---|---|---|---|---|
| Mann-Whitney (medians) | 99 vs. $80^{***}$ | 69 vs. $60^{***}$ | .579 vs. $.563^{***}$ | .313 vs. $.246^{***}$ |
| OLS | $.185^{***}$ (.018) | $.044^{*}$ (.021) | $.030^{***}$ (.003) | $.141^{***}$ (.013) |
| PSM ATT | .251 | .143 | .030 | .108 |
| Cross-fit AIPW ATE | $.145^{***}$ (.026) | .005 (.031) | $.031^{***}$ (.004) | $.140^{***}$ (.017) |

*Note.* OLS/AIPW: estimate (SE); PSM: ATT. Log outcomes share 3,601 matched pairs; reply share uses 3,600 separately matched pairs (Appendix D.2). $^{*}p < .05$; $^{***}p < .001$.

documentation, testimony, visual evidence, or closely related material. Of these, 3,601 posts, or 14.4%, carry an explicit ordinary-person submission marker and constitute the user-submitted witness posts. The remaining 21,492 witness-relevant posts form the primary comparison group. Within that comparison group, 2,083 posts are attributed to institutional, media, or official sources, yielding 5,684 observations in the focused formal-source comparison. The full sample composition and analytic flow are reported in Appendix A.3.

## 4.2. User-Submitted Posts Receive More Replies Than Other Witness-Relevant Posts

User-submitted witness posts receive more replies than other witness-relevant posts. Median reply counts are 99 and 80, respectively (Mann–Whitney $U$ test, $p < .001$). The OLS contrast is .185 on the log(1 + replies) scale ($SE$ = .018), corresponding to a 20.3% difference in the geometric mean of 1 + replies. The matched ATT point estimate is .251 (28.5%), while cross-fit AIPW yields a smaller ATE of .145 ($SE$ = .026; 15.6%, $p < .001$). H1 is supported (see Table 1).

## 4.3. User-Submitted Posts Show a More Reply-Oriented Profile

The AIPW retweet contrast is near zero, whereas reply share and the reply–retweet log ratio both indicate a more reply-oriented profile for user-submitted posts than for other witness-relevant posts. User-submitted posts have a median of 69 retweets, compared with 60 for other witness-relevant posts. The AIPW contrast is .005 on the log(1 + retweets) scale, corresponding to an approximately 0.5% difference in the geometric mean of 1 + retweets ($p$ = .88).

Median reply share of reply–retweet activity is .579 for user-submitted posts and .563 for other witness-relevant posts. Median reply–retweet log ratios are .313 and .246, respectively (both $p < .001$). PSM contrasts are .108 for the log ratio and .030 for reply share, using separate matches. AIPW contrasts are .140 and .031, respectively (both $p < .001$).

The positive adjusted contrasts in both relative measures support H2 (see Table 1).

## 4.4. Audited Replies Predominantly Engage the Focal Witness Material

Replies to the audited user-submitted posts predominantly engage the focal witness material. Of 360 replies sampled equally across 72 posts, 329, or 91.4%, constitute substantive uptake. The proportion remains stable across reply-volume strata at 90.8% for low-reply posts and 91.7% for both medium- and high-reply posts. Intercoder agreement reaches 99.2%, with Cohen's $\kappa$ = .95 (see Appendix C.2).

Illustrative replies, paraphrased without usernames, add a local lockdown timeline (informational supplementation), question an official casualty count by invoking incident-grading rules (skepticism), and compare an officer's identity with an official commendation page (external checking). Whether reply composition differs for other witness-relevant posts remains untested.

## 4.5. Higher Reply Volume Persists Across Operational Definitions and Comparison Groups

The association with higher reply volume persists across operational definitions of ordinary-person source attribution, but the formal-source comparison provides less consistent evidence for a more reply-oriented profile. Within the full witness-relevant sample, all three operational definitions (source-coded, literal-marker, and expanded-marker) produce positive AIPW estimates for reply volume, reply share, and the reply–retweet log ratio. Contrasts in geometric means of 1 + replies range from 15.6% to 20.7%, reply-share estimates from .023 to .031, and log-ratio estimates from .091 to .140 (all $p < .001$).

The stricter formal-source comparison retains a positive reply-volume estimate under each definition,

although precision varies. Under the primary source-coded definition, the adjusted reply-volume contrast against institutional, media, or official sources is positive ($b$ = .164 on the log(1 + replies) scale, $p$ = .003). The expanded-marker estimate is 38.4% ($p$ = .003), while the literal-marker estimate is 20.7% with limited precision ($p$ = .091). For the primary source-coded definition, the reply–retweet log-ratio contrast is .071 ($p$ = .045), whereas the reply-share contrast is .010 ($p$ = .160) (see Appendix D.3).

## 4.6. The Reply Association Persists Among Semantically Matched Posts

Semantic-content matching retains 620 of the 3,596 user-submitted posts with usable text, yielding 620 matched pairs and a 17.2% match rate. All 620 pairs are matched within the same year-month and media configuration; none requires the year-month-only fallback. Median cosine similarity is .740, with a 10th–90th percentile range of .707–.811. All reported post-match standardized mean differences are below .09.

Within these matched pairs, user-submitted posts have a .091 higher mean value of log(1 + replies) than their matched comparison posts (95% CI [.005, .175]), corresponding to a 9.5% difference in the geometric mean of 1 + replies (95% CI [0.5%, 19.1%]). The matched differences are .022 (95% CI [.010, .035]) for reply share and .105 (95% CI [.048, .163]) for the reply–retweet log ratio. The association therefore remains detectable among posts meeting the semantic common-support criterion, suggesting that it is not entirely attributable to broad semantic differences captured by the encoder. The result is local to closely matchable posts, and differences in event severity, novelty, evidence quality, and other unmeasured content features may remain (see Appendix D.1).

# 5. Discussion

Digital witnessing continues after testimony enters public view through the ways audiences circulate, contextualize, evaluate, supplement, and contest witness claims. Relational accounts locate witnessing in testimony, mediation, and reception, while digital-witnessing research explains how citizen material becomes visible and travels across networks (Kyriakidou, 2015; Mortensen, 2015; Peters, 2001). The findings specify reception as a differentiated part of this process. Witness-relevant posts carrying ordinary-person source attribution receive more replies than other witness-relevant posts, while the primary AIPW retweet estimate is near zero and the balance of response shifts toward replies in the full witness-relevant sample. Post-publication witnessing therefore includes distinct forms of audience uptake whose relation to testimony differs even when they appear together as platform engagement.

Source attribution is associated with how much public response gathers around witness material after publication. Source-cue research has concentrated on credibility judgments, truth assessments, content selection, and sharing intentions (Kaiser et al., 2021; Metzger et al., 2010; Yaqub et al., 2020). The present findings extend that problem to post-publication response. On the Teacher Li account, ordinary-person source attribution identifies situated access within material that has also passed through Li's selection and presentation. Citizen-witness research shows that proximity, personal experience, and local knowledge can give ordinary-person testimony public relevance (Andén-Papadopoulos, 2014; Kampka & Modrzejewska, 2026). The observed reply pattern is consistent with such attribution presenting a claim whose context remains available for additional knowledge, comparison, and evaluation.

The association between ordinary-person source attribution and higher reply volume remains consistent across stricter comparison and matching specifications. The reply difference persists in the comparison with posts attributed to institutional, media, and official sources. Semantic-content matching also retains a positive reply difference among closely matchable posts. Evidence for a broader shift in the balance between replies and retweets is less consistent in the formal-source comparison. Ordinary-person source attribution therefore corresponds more consistently to reply volume than to a general reallocation of engagement from retweeting toward replying.

The reply audit connects that aggregate pattern to what audiences do around witness claims. Of the 360 sampled replies, 91.4% engaged with the focal witness material. The audited responses supplied contextual information, questioned reported details, compared claims with external information, interpreted institutions and political actors, and evaluated the events described. These practices match forms of collaborative interpretation and verification identified in research on networked testimony and conversational threads (Hermida, 2012; Zubiaga et al., 2016). Reply volume can therefore be read, within the audited setting, as a trace of substantive response to witness material.

The findings also specify how a diaspora-based witness account can support counterpublic communication. Counterpublic theory locates alternative public activity in the circulation and interpretation of claims marginalized within dominant information channels (Fraser, 1990; Warner, 2002), and networked platforms carry such claims across dispersed participants (Jackson & Welles, 2015). Teacher Li links these processes through cross-border curation. Reports and observations originating in mainland China enter an overseas Chinese-language political space through Li's selection and publication. Retweets carry those claims into additional networks, and replies gather contextualization, evaluation, and dispute around the same focal events (Zeng & Cheng, 2025). The account's counterpublic relevance lies in this combination of curation, redistribution, and public response. The circulation, contextualization, and contestation observed around these witness claims constitute forms of counterpublic communication, while collective identity, shared political orientation, and durable counterpublic formation involve a broader set of social relations than those examined here. Digital witnessing supplies shared claims about events, and platform interaction keeps those claims circulating and available for continued interpretation beyond the domestic platform environment.

*Limitations.* The observational design leaves unmeasured content and editorial confounding after adjustment and semantic matching; audience perceptions, credibility judgments, and motivations for replying were not measured, so the situated-testimony interpretation is inferred from interaction data rather than tested; the reply audit covers only November 2022 user-submitted posts with at least five retrievable replies, and engagement measures are cumulative snapshots from a single account on X.

## 6. Conclusion

Digital witnessing does not end when testimony becomes visible. On the Teacher Li account, witness-relevant posts carrying ordinary-person source attribution receive more replies than other witness-relevant posts, the primary AIPW retweet estimate is near zero, and audited replies predominantly engage the focal material with added context, skepticism, and evaluation. The association is consistent with ordinary-person attribution presenting material as situated testimony open to supplementation, evaluation, and contestation. For a Chinese-language public assembled beyond domestic censorship, this differentiated reception is where counterpublic communication becomes observable: curation gathers scattered testimony into a shared record, retweets carry witness claims across dispersed networks, and reply threads hold the contextualization and dispute that domestic platforms foreclose. These interaction traces mark counterpublic communication; collective identity and durable counterpublic formation involve social relations beyond the interactions examined here. Within that scope, the study relocates the analytical endpoint of digital witnessing from the visibility of testimony to its public life after publication, and it places visible source attribution among the conditions under which a dispersed public takes marginalized testimony up.

## 7. Notes

1. Chinese: 李老师不是你老师; X: @whyyoutouzhele.
2. The primary marker is 网友投稿 ("netizen submission"); equivalent markers are defined in the Data and Methods.
3. Model identifier: google/gemma-4-31B-it.
4. Model identifier: NousResearch/Hermes-3-Llama-3.1-405B.
5. Codebook label: S1_user_submission.